\documentclass[showpacs,eqsecnum,preprintnumbers]{revtex4}
\usepackage{amsmath}
\usepackage{amsfonts}
\usepackage{amssymb}

\begin{document}

\title{Inequalities for nucleon generalized parton distributions with helicity flip}
\author{M. Kirch${}^{1}$, P.V. Pobylitsa${}^{1,2}$, and K. Goeke${}^{1}$}
\affiliation{${}^{1}$Institute for Theoretical Physics II, Ruhr University Bochum, D-44780, Germany\\
${}^{2}$Petersburg Nuclear Physics Institute, Gatchina, St.~Petersburg,
188300, Russia}

\begin{abstract}
Several positivity bounds are derived for generalized parton distributions
(GPDs) with helicity flip.
\end{abstract}
\maketitle

\section{Introduction}

\subsection{Generalized parton distributions}

The theoretical description of strong interactions in hard processes is based
on the concept of QCD factorization. In many hard processes the quark-gluon
interaction at large momentum transfer can be described perturbatively,
whereas the quark-gluon structure of the initial and final hadrons still
remains nonperturbative. Sometimes this physical idea can be put on solid
theoretical ground. The so-called factorization theorems represent observable
physical cross sections or amplitudes in terms of convolutions of
perturbatively computable coefficient functions and nonperturbative objects
like parton distributions, fragmentation functions, distribution amplitudes,
generalized parton distributions, etc.

The rich realm of factorization theorems can be divided in two different groups:

1) factorization of cross sections [usually associated with inclusive
(semi-inclusive) processes],

2) factorization of amplitudes in exclusive reactions.

The border between the two groups is not absolute. The basic principles of
quantum theory include the concept of the density matrix which allows one to
combine the description of probabilities with the physics of interference.
Another connection between the two classes of factorization theorems is
provided by the optical theorem which gives us a relation between amplitudes
and cross sections.

A classical example of factorization for cross sections is deeply inelastic
scattering where the required information on nonperturbative physics can
be condensed into \emph{parton distributions}. In the second group of hard
processes with the factorization theorems for amplitudes an important role is
played by the deeply virtual Compton scattering (DVCS) which can be described
in terms of the \emph{generalized parton distributions} (GPDs)
\cite{MRGDH-94,Radyushkin-96-a,Radyushkin-96,Ji-97,Ji-97-b,CFS-97,Radyushkin-97,Radyushkin-review,GPV,BMK-2001,AMS,Diehl-2003,Belitsky:2005qn}.

In spite of the different physical meaning of usual parton distributions
(probabilities) and GPDs (amplitudes), their theoretical description is based
on the matrix elements of the same bilocal light-ray operators. For example,
in the case of quark distributions one deals with the operators
\begin{equation}
O_{\Gamma}(\lambda,n)=\bar{\psi}\left(  -\frac{\lambda n}{2}\right)
\Gamma\psi\left(  \frac{\lambda n}{2}\right)  \,, \label{light-ray-ME}
\end{equation}
where $\Gamma$ is some spin matrix and the quark field $\psi$ is taken along
the light-like vector $n$
\begin{equation}
n^{2}=0\,.
\end{equation}
The GPDs are defined in terms of \emph{nondiagonal} matrix elements of the
bilocal operator (\ref{light-ray-ME})
\begin{equation}
\int\frac{d\lambda}{2\pi}e^{i\lambda x}\langle P_2|O_{\Gamma}
(\lambda,n)|P_1\rangle\quad\mathrm{(GPDs)}\,, \label{ME-GPD-1}
\end{equation}
whereas the definition of usual parton distributions is based on
\emph{diagonal} matrix elements of the same operator. These diagonal matrix
elements can be obtained by taking the limit $P_2\rightarrow P_1$ in Eq.
(\ref{ME-GPD-1}). Therefore the usual parton distributions are sometimes
called \emph{forward parton distributions} (FPDs). Thus, FPDs can be expressed
in terms of the matrix elements
\begin{equation}
\int\frac{d\lambda}{2\pi}e^{i\lambda x}\langle P_1|
O_{\Gamma}(\lambda,n)|P_1\rangle\quad\mathrm{(FPDs)\,}\,.
\end{equation}

In this paper we will concentrate on the so-called helicity-flip GPDs. These
GPDs correspond to operators $O_{\Gamma}$ with matrices $\Gamma$ changing the
helicity of the parton. Helicity-flip GPDs were introduced by Hoodbhoy and Ji in Ref.
\cite{HJ-1998}. Their complete classification (for the twist-two case) was given
by Diehl in Ref. \cite{Diehl:2001pm}. Phenomenological applications of the
helicity-flip GPDs are discussed in Refs.
\cite{HJ-1998,Diehl:2001pm,BMK-2001,IPST-02,IPST-03,IPST-04}.

\subsection{Positivity bounds on GPDs}

Now we want to describe the main idea standing behind the derivation of the
positivity bounds for GPDs. The precise classification of GPDs (which is
discussed in Sec. \ref{GPD-classification-subsection}) is not important for
this brief derivation.

Let us consider the states
\begin{equation}
|x\sigma PS\rangle=\int\frac{d\lambda}{2\pi}e^{ix\lambda}\psi_{\sigma}\left(
\frac{\lambda n}{2}\right)  |PS\rangle\,,
\end{equation}
where $\sigma$ and $S$ stand for the quark and hadron polarizations.

Using the positivity of the norm in the Hilbert space of states
\begin{equation}
\langle c|c\rangle\geq0
\end{equation}
for the linear combinations
\begin{equation}
|c\rangle=\sum\limits_{m}c_{m}|x_{m}\sigma_{m}P_{m}S_{m}\rangle\,,
\label{c-combination}
\end{equation}
we find
\begin{equation}
\langle c|c\rangle=\sum\limits_{km}c_{k}^{\ast}c_{m}\langle x_{k}\sigma
_{k}P_{k}S_{k}|x_{m}\sigma_{m}P_{m}S_{m}\rangle\geq0\,. \label{c-Pi-c}
\end{equation}
This inequality can be rewritten in the form
\begin{equation}
-\sum\limits_{k\neq m}c_{k}^{\ast}c_{m}\langle x_{k}\sigma_{k}P_{k}S_{k}
|x_{m}\sigma_{m}P_{m}S_{m}\rangle\leq\sum\limits_{k}\left|  c_{k}\right|
^{2}\langle x_{k}\sigma_{k}P_{k}S_{k}|x_{k}\sigma_{k}P_{k}S_{k}\rangle\,.
\label{cP-cP}
\end{equation}
The nondiagonal matrix elements on the left-hand side (LHS) are associated with GPDs whereas
the diagonal matrix elements of the right-hand side (RHS)
correspond to FPDs. Using the freedom
of choice of arbitrary coefficients $c_{k}$, one can derive positivity bounds
on GPDs from inequality (\ref{cP-cP}):
\begin{equation}
|\mathrm{GPD}|\leq f\left(  \mathrm{FPD}\right)  \,, \label{ineq-structure}
\end{equation}
where function $f$ depends on the set of coefficients $c_{i}$ used in the
derivation of the inequality.

Various inequalities for GPDs have been derived in Refs.~\cite{Martin-98,
Teryaev-98a,Teryaev-98b,Radyushkin-99,PST-99,Ji-98,DFJK-00,Burkardt-01,Pobylitsa-01,Pobylitsa-02,Diehl-02,
Burkardt-02-a,Burkardt-02-b,Burkardt:2003ck,Diehl-Hagler-05}. As shown in
Ref.~\cite{Pobylitsa-02-c}, these inequalities can be considered as particular
cases of a general positivity bound which has a relatively simple form in the
impact parameter representation for GPDs
\cite{Burkardt-01,Diehl-02,Burkardt-02-a,Burkardt-02-b,Burkardt-00,Burkardt:2003ck}.
In the brief derivation of positivity bounds sketched above, we ignored
the problems of the light-cone singularities, renormalization, gauge invariance, etc.
These issues are discussed in Refs. \cite{Pobylitsa-02-c,Pobylitsa-04}.
The positivity bounds are stable under the one-loop evolution of GPDs to
higher normalization points \cite{Pobylitsa-02-c}. The positivity bound of
Ref.~\cite{Pobylitsa-02-c} was explicitly checked for one-loop GPDs in various
perturbative models \cite{Pobylitsa-02-e}. The solutions of the combined
positivity and polynomiality constraints
are studied in Refs.~\cite{Pobylitsa-02-d,Pobylitsa-02-e}. The
positivity bounds can be used for self-consistency checks of models of GPDs
\cite{TM-02,Mukherjee:2002gb,Tiburzi:2002tq,Tiburzi-04,TDM-04,CM-2005}.

\subsection{Standard classification of twist-two quark GPDs}

\label{GPD-classification-subsection}

The standard classification of the leading-twist quark FPDs includes three distributions
\begin{equation}
\mathrm{FPDs:}\quad q,\Delta_{L}q,\Delta_{T}q \label{FPDs}
\end{equation}
known as unpolarized, polarized and transversity, respectively. Any of these
FPDs can be represented as the forward limit of some GPD but the inverse
statement is not true. For some GPDs the forward limit cannot be associated
with FPDs. Therefore the number of leading-twist GPDs is larger than the
number of FPDs. The full set of the leading-twist quark GPDs includes eight
functions \cite{Diehl:2001pm}
\begin{equation}
\mathrm{GPDs:\quad}H^{q},E^{q},\tilde{H}^{q},\tilde{E}^{q},H_{T}^{q},E_{T}
^{q},\tilde{H}_{T}^{q},\tilde{E}_{T}^{q}\,. \label{twist-2-GPDs}
\end{equation}

The twist-two operators correspond to the following constraint on the matrix
$\Gamma$ appearing in Eq. (\ref{light-ray-ME})
\begin{equation}
\Gamma(n\gamma)=(n\gamma)\Gamma=0\,. \label{Gamma-constraint}
\end{equation}
The solutions of this constraint are
\begin{equation}
\Gamma=(n\gamma),\,(n\gamma)\gamma_{5},\,[(a\gamma),(n\gamma)]\,,
\label{Gamma-cases}
\end{equation}
where vector $a$ should obey the condition
\begin{equation}
(an)=0\,. \label{a-n-0}
\end{equation}
The three cases (\ref{Gamma-cases}) correspond to three usual FPDs (\ref{FPDs}).

The set of GPDs associated with the same light-ray operators is larger. The
covariant decomposition of the nucleon matrix elements of
operators (\ref{light-ray-ME})
leads to eight twist-two parton GPDs (\ref{twist-2-GPDs}).
These GPDs are listed in Table \ref{table1}.

\begin{table}[h]
\caption{\label{table1}
Twist-two quark GPDs and FPDs}
\begin{tabular}
[c]{|l|l|l|}\hline
$\Gamma$ & GPDs & FPDs\\\hline
$(n\gamma)$ & $H^{q},E^{q}$ & $q$\\\hline
$(n\gamma)\gamma_{5}$ & $\tilde{H}^{q},\tilde{E}^{q}$ & $\Delta_{L}q$\\\hline
$\lbrack(a\gamma),(n\gamma)]$ & $H_{T}^{q},E_{T}^{q},\tilde{H}_{T}^{q}
,\tilde{E}_{T}^{q}$ & $\Delta_{T}q$\\\hline
\end{tabular}
\end{table}

We assume the normalization of the light-like vector $n$
\begin{equation}
n(P_{1}+P_{2})=2
\end{equation}
and the standard notation
\begin{equation}
\Delta=P_{2}-P_{1}\,,
\end{equation}
\begin{equation}
\bar{P}=\frac{1}{2}(P_{1}+P_{2})\,.
\end{equation}
Then the precise definitions of quark GPDs are
\begin{align}
&  \int\frac{d\lambda}{2\pi}e^{i\lambda x}\langle U(P_{2})|\bar{\psi}\left(
-\frac{\lambda n}{2}\right)  (n\gamma)\psi\left(  \frac{\lambda n}{2}\right)
|U(P_{1})\rangle\nonumber\\
&  =\bar{U}(P_{2})\left[  H^{q}(n\gamma)+\frac{1}{2m}E^{q}i\sigma^{\mu\nu
}n_{\mu}\Delta_{\nu}\right]  U(P_{1})\,, \label{H-E-def}
\end{align}
\begin{align}
&  \int\frac{d\lambda}{2\pi}e^{i\lambda x}\langle U(P_{2})|\bar{\psi}\left(
-\frac{\lambda n}{2}\right)  (n\gamma)\gamma_{5}\psi\left(  \frac{\lambda
n}{2}\right)  |U(P_{1})\rangle\nonumber\\
&  =\bar{U}(P_{2})\left[  \tilde{H}^{q}(n\gamma)\gamma_{5}+\frac{1}{2m}
\tilde{E}^{q}\gamma_{5}(n\Delta)\right]  U(P_{1})\,,
\end{align}
\begin{align}
&  \int\frac{d\lambda}{2\pi}e^{i\lambda x}\langle U(P_{2})|\frac{1}{2}
\bar{\psi}\left(  -\frac{\lambda n}{2}\right)  \left[  (a\gamma),(n\gamma
)\right]  \psi\left(  \frac{\lambda n}{2}\right)  |U(P_{1})\rangle\nonumber\\
&  =\bar{U}(P_{2})\left\{  \frac{1}{2}H_{T}^{q}\left[  (a\gamma),(n\gamma
)\right]  +\tilde{H}_{T}^{q}\frac{(\bar{P}n)(\Delta a)-(\Delta n)(\bar{P}
a)}{m^{2}}\right. \nonumber\\
&  \left.  +E_{T}^{q}\frac{(\gamma n)(\Delta a)-(\Delta n)(\gamma a)}
{2m}+\tilde{E}_{T}^{q}\frac{(\gamma n)(\bar{P}a)-(\bar{P}n)(\gamma a)}
{m}\right\}  U(P_{1})\,. \label{T-GPDs-def}
\end{align}
Here $m$ is the nucleon mass, and $a$ is an arbitrary vector obeying
condition (\ref{a-n-0}).

The GPDs depend on the standard Ji variables \cite{Ji-98} $x$ [which appears
in the exponent $e^{i\lambda x}$ on the LHS of 
Eqs. (\ref{H-E-def})--(\ref{T-GPDs-def})] and $\xi,t$:
\begin{equation}
\xi=\frac{n(P_{1}-P_{2})}{n(P_{1}+P_{2})}\,,\quad t=(P_{2}-P_{1})^{2}
=\Delta^{2}\,. \label{xi-t-def}
\end{equation}
At fixed $\xi$ the allowed region for the variable $t$ is
\begin{equation}
t<t_{0}<0\,,
\end{equation}
where
\begin{equation}
t_{0}=-\frac{4m^{2}\xi^{2}}{1-\xi^{2}}\,. \label{t0-def}
\end{equation}

\subsection{Compact notation for kinematic variables}

Most of the positivity bounds for GPDs look rather cumbersome if one writes
them directly in terms of the standard Ji variables $x,\xi,t$. Both the
derivation and the final representation of the positivity bounds become much
simpler if one introduces several new variables.

The first pair of auxiliary variables
\begin{equation}
x_{1}=\frac{x+\xi}{1+\xi}\,,\quad x_{2}=\frac{x-\xi}{1-\xi} \label{x-k-def}
\end{equation}
is well known. The variable $x_{1}$ ($x_{2}$) has the meaning of the momentum
fraction of the initial (final) quark with respect to the initial (final)
nucleon. The positivity bounds are derived for the region
\begin{equation}
|\xi|<x<1\,, \label{x-xi-quarks}
\end{equation}
where the variables $x_{k}$ obey the constraint
\begin{equation}
0<x_{k}<1\,. \label{x-k-physical}
\end{equation}
We also define the variables
\begin{equation}
\alpha=\frac{1}{\sqrt{1-\xi^{2}}}\,, \label{alpha-beta-def-0}
\end{equation}
\begin{equation}
\nu=\frac{2m}{\sqrt{\left(  t_{0}-t\right)  \left(  1-\xi^{2}\right)  }
}=\frac{1}{\xi}\sqrt{\frac{-t_{0}}{t_{0}-t}}\,. \label{gamma-def}
\end{equation}
In terms of these variables we have
\begin{equation}
t_{0}=-4\alpha^{2}m^{2}\xi^{2}\,,
\end{equation}
\begin{equation}
t=-4\alpha^{2}m^{2}\left(  \nu^{-2}+\xi^{2}\right)  \,, \label{t-nu-xi}
\end{equation}
\begin{equation}
\left|  \frac{t_{0}}{t_{0}-t}\right|  =\frac{-t_{0}}{t_{0}-t}=\nu^{2}\xi
^{2}\,, \label{gamma-2-xi-2}
\end{equation}
\begin{equation}
\frac{m^{2}}{t_{0}-t}=\frac{\nu^{2}}{4\alpha^{2}}\,.
\end{equation}

\subsection{Short guide through positivity bounds}

The literature on the positivity bounds for GPDs is rather extensive. Without
trying to give a complete review of the accumulated results, let us make
several comments which may be helpful for the orientation in the realm of
inequalities for GPDs.

Starting from the general inequality (\ref{c-Pi-c}), one can reduce this
inequality to others using various strategies:

1) One can concentrate on the case when the linear combination
(\ref{c-combination}) contains only two configurations for the quark-hadron
kinematics ($x_{1}$, $P_{1}$ and $x_{2}$, $P_{2}$) but includes all possible
quark and nucleon polarizations. This way is chosen in this paper.

2) One can include all possible values of $x_m$, $P_m$ in the linear
combination (\ref{c-combination}) but concentrate on the minimal set of
quark and nucleon polarizations.
This leads to the positivity bounds in the impact parameter
representation which were derived in Ref. \cite{Pobylitsa-02-c} for
helicity-nonflip GPDs.

3) One can choose an intermediate way by taking the parton-hadron
configurations $x$, $P$ with fixed $x$ (but arbitrary $P$) and work with
arbitrary polarizations. This leads to the positivity bounds for GPDs with
$\xi=0$. The case of helicity-flip GPDs was considered in this context in Ref.
\cite{Diehl-Hagler-05}.

\section{Examples of the results}

\label{Examples-section}

\subsection{Simple inequality for GPD $H_{T}^{q}$}

As was already mentioned, our aim is to derive bounds on helicity-flip GPDs
which have the structure (\ref{ineq-structure}). In order to give a
preliminary impression about the results obtained in this paper let us
consider one example:
\begin{equation}
|H_{T}^{q}(x,\xi,t)|\leq\sqrt{q\left(  \frac{x+\xi}{1+\xi}\right)  q\left(
\frac{x-\xi}{1-\xi}\right)  \left|  (1-\xi^{2})+\frac{4m^{2}\xi^{2}}
{t}\right|  ^{-1}}\,. \label{H-T-example}
\end{equation}
On the LHS we deal with the helicity-flip GPD $H_{T}^{q}$ (\ref{T-GPDs-def})
depending on the Ji variables $x,\xi,t$ (\ref{xi-t-def}). On the RHS we have
the product of two unpolarized quark FPDs $q$ taken at values $x_{k}$
(\ref{x-k-def}):
\begin{equation}
q_{k}=q(x_{k})\,. \label{q-k-def}
\end{equation}
This positivity bound (as well as other positivity bounds) can be used in the
domain (\ref{x-xi-quarks}), where the variables $x_{k}$ belong to the physical
region $0<x_{k}<1$ (\ref{x-k-physical}) and the FPDs $q(x_{k})$ do not
vanish. In terms of variables $\alpha$ (\ref{alpha-beta-def-0}), $\nu$
(\ref{gamma-def}) and $q_{k}$ (\ref{q-k-def}) we can rewrite inequality
(\ref{H-T-example}) in the compact form
\begin{equation}
|H_{T}^{q}|\leq\alpha\sqrt{\left(  1+\nu^{2}\xi^{2}\right)  q_{1}q_{2}}\,.
\label{H-T-example-1b}
\end{equation}
Note that this positivity bound for the GPD $H_{T}^{q}$ contains only the
unpolarized FPD $q$.

\subsection{Enhanced positivity bound on $H_{T}^{q}$}

\label{Enhanced-HTq-section}

In fact, using the polarized FPD\ $\Delta_{L}q$ in addition to the unpolarized
FPD $q$, one can derive another bound for the same GPD:
\begin{equation}
|H_{T}^{q}|\leq\frac{\alpha}{2}\left[  \left(  r_{1}^{-}r_{2}^{-}+r_{1}
^{+}r_{2}^{+}\right)  ^{2}+\nu^{2}\xi^{2}\left(  r_{1}^{+}r_{2}^{-}+r_{1}
^{-}r_{2}^{+}\right)  ^{2}\right]  ^{1/2}\,. \label{H-T-example-2}
\end{equation}
Here we use the compact notation
\begin{equation}
r_{k}^{\pm}=\sqrt{q(x_{k})\pm\Delta_{L}q(x_{k})}\,. \label{r-k-pm-def}
\end{equation}
Note that the square root on the RHS is applied to a positive expression since
\begin{equation}
|\Delta_{L}q|\leq q\,. \label{Delta-L-region}
\end{equation}

Let us show that the previous inequality (\ref{H-T-example-1b}) is a trivial
consequence of the stronger inequality (\ref{H-T-example-2}). Indeed, starting
from the general inequality
\begin{equation}
\left|  a_{1}b_{1}+a_{2}b_{2}\right|  =|\mathbf{a}\cdot\mathbf{b}
|\leq|\mathbf{a}||\mathbf{b}|=\sqrt{\left(  a_{1}\right)  ^{2}+\left(
a_{2}\right)  ^{2}}\sqrt{\left(  b_{1}\right)  ^{2}+\left(  b_{2}\right)
^{2}}\,,
\end{equation}
we find
\begin{align}
 r_{1}^{-}r_{2}^{-}+r_{1}^{+}r_{2}^{+}   &  \leq
\sqrt{\left(  r_{1}^{-}\right)  ^{2}+\left(  r_{1}^{+}\right)  ^{2}}
\sqrt{\left(  r_{2}^{-}\right)  ^{2}+\left(  r_{2}^{+}\right)  ^{2}
}\,,\label{rr-ineq-1}\\
  r_{1}^{+}r_{2}^{-}+r_{1}^{-}r_{2}^{+} &  \leq
\sqrt{\left(  r_{1}^{-}\right)  ^{2}+\left(  r_{1}^{+}\right)  ^{2}}
\sqrt{\left(  r_{2}^{-}\right)  ^{2}+\left(  r_{2}^{+}\right)  ^{2}}\,.
\label{rr-ineq-2}
\end{align}
According to Eq. (\ref{r-k-pm-def}) we have
\begin{equation}
\left(  r_{k}^{-}\right)  ^{2}+\left(  r_{k}^{+}\right)  ^{2}=2q_{k}\,.
\label{r2-r2-q}
\end{equation}
Taking a linear combination of inequalities (\ref{rr-ineq-1}),
(\ref{rr-ineq-2}) and using Eq. (\ref{r2-r2-q}), we find
\begin{equation}
\left(  r_{1}^{-}r_{2}^{-}+r_{1}^{+}r_{2}^{+}\right)  ^{2}+\nu^{2}\xi
^{2}\left(  r_{1}^{+}r_{2}^{-}+r_{1}^{-}r_{2}^{+}\right)  ^{2}\leq4\left(
1+\nu^{2}\xi^{2}\right)  q_{1}q_{2}\,.
\end{equation}
Using this inequality, we see that positivity bound (\ref{H-T-example-1b}) is a
consequence of inequality (\ref{H-T-example-2}).

\subsection{Positivity bound on $H_{T}^{q}$ using transversity FPD}

\label{H-stong-bound-example-section}

Inequality (\ref{H-T-example-2}) can be also enhanced. In order to write the
enhanced positivity bound in a simple form we introduce a special notation.
Starting from FPDs (\ref{FPDs}), let us construct the following combinations
\begin{equation}
\left(
\begin{array}
[c]{c}
Q^{1}\\
Q^{2}\\
Q^{3}\\
Q^{4}
\end{array}
\right)  =\left(
\begin{array}
[c]{c}
q+\Delta_{L}q+2\Delta_{T}q\\
q-\Delta_{L}q\\
q+\Delta_{L}q-2\Delta_{T}q\\
q-\Delta_{L}q
\end{array}
\right)  \,. \label{Q1-4-def}
\end{equation}
One entry is duplicated here:
\begin{equation}
Q^{2}=Q^{4} \label{Q-4l-Q-2l}
\end{equation}
but this redundancy will be helpful for our later work in Sec.
\ref{General-bounds-section}.

Using Eq. (\ref{Delta-L-region}) and the Soffer inequality \cite{Soffer-95}
\begin{equation}
2|\Delta_{T}q|\leq q+\Delta_{L}q\,, \label{Soffer-ineq}
\end{equation}
we conclude that all functions $Q^{M}$ are positive:
\begin{equation}
Q^{M}\geq0\,.
\end{equation}
It is convenient to introduce the notation
\begin{equation}
R_{k}^{M}=\sqrt{Q^{M}(x_{k})} \label{R-jk-def}
\end{equation}
with $x_{k}$ defined by Eq. (\ref{x-k-def}).

Then our inequality for $H_{T}^{q}$ has the form
\begin{equation}
|H_{T}^{q}|\leq\frac{\alpha}{4}\sum\limits_{k=1,3}\left[  \left(  R_{1}
^{k}R_{2}^{k}+R_{1}^{2}R_{2}^{2}\right)  ^{2}+\nu^{2}\xi^{2}\left(  R_{1}
^{k}R_{2}^{2}+R_{1}^{2}R_{2}^{k}\right)  ^{2}\right]  ^{1/2}\,.
\label{H-T-example-3}
\end{equation}
In Appendix \ref{Transversity-exclusion-appendix} we show that the previous
positivity bound (\ref{H-T-example-2}) is an algebraic consequence of
inequality (\ref{H-T-example-3}).

Let us summarize. We have presented three inequalities (\ref{H-T-example}),
(\ref{H-T-example-2}) and (\ref{H-T-example-3}) for $H_{T}^{q}$. Inequality
(\ref{H-T-example-3}) is the strongest one, it can be used as a starting point
for the derivation of inequality (\ref{H-T-example-2}). In turn, positivity
bound (\ref{H-T-example}) follows from Eq. (\ref{H-T-example-2}). Thus, the
problem reduces to the derivation of the strongest bound (\ref{H-T-example-3}).
We postpone this derivation till Sec. \ref{Example-H-q-T-section}.

Although formally the bound (\ref{H-T-example-3}) is the strongest one, in
practical applications it is not too helpful since it contains the
transversity distribution $\Delta_{T}q$ which is poorly known. In Appendix
\ref{Transversity-exclusion-appendix} we show how the maximization of the RHS
of the bound (\ref{H-T-example-3}) with respect to the unknown transversity
distribution $\Delta_{T}q$ leads to the weaker inequality (\ref{H-T-example-2}).
This weaker bound contains the forward distributions $q$ and $\Delta_{L}q$
on the RHS. Depending on the availability of phenomenological data on
$\Delta_{L}q$ one can either use this bound directly or one can maximize its
RHS with respect to $\Delta_{L}q$. In the second case one arrives at the
weakest bound (\ref{H-T-example}) for $H_{T}^{q}$.

We have described the hierarchy of positivity bounds the GPD $H_{T}^{q}$.
The case of other GPDs is similar.
Below we will derive the bounds of the strongest type (\ref{H-T-example-3})
for all helicity-flip GPDs. After that we will show how weaker bounds can be
derived by eliminating $\Delta_{T}q$ and $\Delta_{L}q$.

\section{General positivity bounds for GPDs}

\label{General-bounds-section}

\subsection{Quark case}

In Sec. \ref{Examples-section} we have considered several examples of
positivity bounds. Now we want to turn to the derivation of the inequalities.
The idea of this derivation is represented by 
Eqs. (\ref{c-Pi-c})--(\ref{ineq-structure}).
As for the technical realization of this idea, an essential part
of the work has already been done in Ref.~\cite{Pobylitsa-02},
where a general positivity bound was derived for the so-called
helicity amplitudes.                
In this section we briefly describe the main result of Ref.
\cite{Pobylitsa-02}. In Sec. \ref{L-G-matrices} we will turn to the analysis
of the special case of positivity bounds for the helicity-flip GPDs.

The general positivity bounds on GPDs were derived in Ref.
\cite{Pobylitsa-02} using a special matrix notation for GPDs and FPDs:

1) matrices $F_{s}^{q}(x)$ made of FPDs $q,\Delta_{L}q,\Delta_{T}q$,

2) matrices $B_{s}^{q}(x,\xi,t)$ made of GPDs $H^{q},E^{q},\tilde{H}
^{q},\tilde{E}^{q},H_{T}^{q},E_{T}^{q},\tilde{H}_{T}^{q},\tilde{E}_{T}^{q}$
(\ref{twist-2-GPDs}).

Both $F_{s}^{q}$ and $B_{s}^{q}$ are $2\times2$ matrices. In addition to the
matrix structure, $F_{s}^{q}$ and $B_{s}^{q}$ have the subscript $s$ which runs
over the values $s=1,2$. The precise definition of $F_{s}^{q}$ and $B_{s}^{q}$
will be discussed later.

Using these GPD matrices $B_{s}^{q}(x,\xi,t)$ and FPD matrices $F_{s}^{q}(x)$,
one can write the general positivity bound derived in Ref. \cite{Pobylitsa-02}
in the form
\begin{equation}
\left|  \sum\limits_{s=1}^{2}\mathrm{Tr}\left[  L_{s}B_{s}^{q}(x,\xi
,t)\right]  \right|  \leq\sum\limits_{s=1}^{2}\mathrm{Tr}\left(  \left[
F_{s}^{q}(x_{1})L_{s}F_{s}^{q}(x_{2})L_{s}^{\dagger}\right]  ^{1/2}\right)
\,. \label{G-bound-general}
\end{equation}
This inequality holds for arbitrary $2\times2$
matrices $L_{s}$ ($s=1,2$).
$L_{s}^{\dagger}$ stands for the matrix Hermitian conjugate to
$L_{s}$.
The power $1/2$ on the RHS of inequality
(\ref{G-bound-general}) should be understood in the matrix sense. The
variables $x_{k}$ ($k=1,2$) are given by Eq. (\ref{x-k-def}).

Using the freedom of choice of matrices $L_{s}$, one can derive bounds for
arbitrary GPDs from the inequality (\ref{G-bound-general}).
In Ref. \cite{Pobylitsa-02} the
general positivity bound (\ref{G-bound-general}) was used for the derivation
of inequalities for the GPDs $H^{q},E^{q},\tilde{H}^{q},\tilde{E}^{q}$. In
this paper we will derive inequalities for the helicity-flip GPDs
$H_{T}^{q},E_{T}^{q},\tilde{H}_{T}^{q},\tilde{E}_{T}^{q}$ from the same general
inequality (\ref{G-bound-general}).

Now we turn to the definition of matrices $F_{s}^{q}$ and $B_{s}^{q}$ which
appear in the general positivity bound (\ref{G-bound-general}). Matrices
$F_{s}^{q}(x)$ are defined via the linear combinations  $Q^{M}(x)$ of FPDs
(\ref{Q1-4-def}):
\begin{equation}
F_{1}^{q}(x)=\left(
\begin{array}
[c]{cc}
Q^{1}(x) & 0\\
0 & Q^{2}(x)
\end{array}
\right)  \,\,,\quad F_{2}^{q}(x)=\left(
\begin{array}
[c]{cc}
Q^{3}(x) & 0\\
0 & Q^{4}(x)
\end{array}
\right)  \,.\label{F1}
\end{equation}
The definition of matrices $B_{s}^{q}(x,\xi,t)$ can be found in Appendix
\ref{A-B-appendix}. In this section we do not need the
explicit expression for $B_{s}^{q}(x,\xi,t)$. But it is important to
understand that the two matrices $B_{s}^{q}$ (with $s=1,2$) contain altogether
$4+4=8$ linearly independent matrix elements. This is exactly the number of
the full set of twist-two GPDs (\ref{twist-2-GPDs}). Hence the matrix elements
$B_{s}^{ij}$ of matrices $B_{s}$ can be used as an alternative basis in the
space of linear combinations of GPDs. Thus, any GPD (and any linear combination
of GPDs) $G$ can be represented in the form
\begin{equation}
G^{q}(x,\xi,t)=\sum\limits_{i,j,s=1}^{2}L_{s}^{ji}(G^{q})\left(  B_{s}
^{q}\right)  ^{ij}(x,\xi,t)\,.
\end{equation}
where $L_{s}^{ji}(G^q)$ are coefficients of the decomposition. Interpreting the
coefficients $L_{s}^{ij}(G^{q})$ as matrix elements of $2\times2$ matrices
$L_{s}(G^{q})$ we can write
\begin{equation}
G^{q}(x,\xi,t)=\sum\limits_{s=1}^{2}\mathrm{Tr}\left[  L_{s}(G^{q})B_{s}
^{q}(x,\xi,t)\right]  \,.\label{G-L-B}
\end{equation}
Inserting this decomposition into the LHS of inequality (\ref{G-bound-general}),
we find
\begin{equation}
\left|  G^{q}(x,\xi,t)\right|  \leq\sum\limits_{s=1}^{2}\mathrm{Tr}\left\{
\left[  F_{s}^{q}(x_{1})L_{s}(G^{q})F_{s}^{q}(x_{2})L_{s}^{\dagger}
(G^{q})\right]  ^{1/2}\right\}  \,.\label{G-bound-general-1A}
\end{equation}

In order to compute the RHS of inequality (\ref{G-bound-general}), we use the
general formula
\begin{equation}
\mathrm{Tr}\,(S^{1/2})=\sqrt{\mathrm{Tr}S+2\sqrt{\det S}}
\label{Tr-sqrt-matrix}
\end{equation}
valid for $2\times2$ matrices $S$. It follows from the fact that any matrix $T$
obeys its characteristic equation
\begin{equation}
P(\lambda)=\det(T-\lambda)\,,\quad P(T)=0\,.
\end{equation}
In the case of $2\times2$ matrices, this results in
\begin{equation}
\det T-(\mathrm{Tr}T)T+T^{2}=0\,. \label{HJ}
\end{equation}
Taking the trace of this equation and substituting $T=S^{1/2}$, we obtain
identity (\ref{Tr-sqrt-matrix}). Using identity (\ref{Tr-sqrt-matrix}) and short
notation $R_{k}^{M}$ (\ref{R-jk-def}), we easily compute the RHS of inequality
(\ref{G-bound-general-1A}):
\begin{align}
|G^{q}|  &  \leq\left\{  \sum\limits_{M,N=1}^{2}\left(  R_{1}^{M}R_{2}
^{N}\right)  ^{2}\left|  \left[  L_{1}(G^{q})\right]  _{MN}\right|
^{2}+2\left|  \det L_{1}(G^{q})\right|  R_{1}^{1}R_{2}^{1}R_{1}^{2}R_{2}
^{2}\right\}  ^{1/2}\nonumber\\
&  +\left\{  \sum\limits_{M,N=1}^{2}\left(  R_{1}^{M+2}R_{2}^{N+2}\right)
^{2}\left|  \left[  L_{2}(G^{q})\right]  _{MN}\right|  ^{2}+2\left|  \det
L_{2}(G^{q})\right|  R_{1}^{3}R_{2}^{3}R_{1}^{4}R_{2}^{4}\right\}  ^{1/2}\,.
\label{G-bound-general-2}
\end{align}

\subsection{Gluon case}

The generalization for the case of gluon GPDs is straightforward.
We follow the definition of gluon GPDs used in Ref. \cite{Diehl:2001pm}.
In the gluon case, inequality (\ref{G-bound-general}) should be modified by
a factor of
$\sqrt{1-\xi^{2}}$ (see Ref. \cite{Pobylitsa-02}):
\begin{equation}
\frac{1}{\sqrt{1-\xi^{2}}}\left|  \sum\limits_{s=1}^{2}\mathrm{Tr}\left[
L_{s}B_{s}^{g}(x,\xi,t)\right]  \right|  \leq\sum\limits_{s=1}^{2}
\mathrm{Tr}\left(  \left[  F_{s}^{g}(x_{1})L_{s}F_{s}^{g}(x_{2})L_{s}
^{\dagger}\right]  ^{1/2}\right)  \,.
\end{equation}
The gluon analogs $F^g_{s}(x)$ of quark matrices (\ref{F1})  are
\begin{equation}
F_{1}^{g}(x)=F_{2}^{g}(x)=\left(
\begin{array}
[c]{cc}
xg^{+}(x) & 0\\
0 & xg^{-}(x)
\end{array}
\right)  ,
\end{equation}
where we use notation
\begin{equation}
g^{\pm}(x)=g(x)\pm\Delta g(x). \label{gpm}
\end{equation}
Here $g(x)$ and $\Delta g(x)$ stand for the forward gluon unpolarized and
polarized distributions, respectively. Note that in the case of spin-$1/2$
hadrons we have no forward gluon transversity distribution.
Thus, the gluon analog of inequality (\ref{G-bound-general-2}) can be obtained
by replacing
\begin{equation}
|G^{q}|\rightarrow\frac{1}{\sqrt{1-\xi^{2}}}|G^{g}|
\end{equation}
on the LHS of inequality (\ref{G-bound-general-2}) and replacing
\begin{equation}
\left(
\begin{array}
[c]{c}
R_{k}^{1}\\
R_{k}^{2}\\
R_{k}^{3}\\
R_{k}^{4}
\end{array}
\right)  \rightarrow\sqrt{x_{k}}\left(
\begin{array}
[c]{c}
y_{k}^{+}\\
y_{k}^{-}\\
y_{k}^{+}\\
y_{k}^{-}
\end{array}
\right)  \,,
\end{equation}
\begin{equation}
y_{k}^{\pm}=\sqrt{g^{\pm}(x_{k})}\,\, \label{s-k-pm-def}
\end{equation}
on the RHS. As a result, we obtain
\begin{equation}
\frac{1}{\sqrt{1-\xi^{2}}}|G^{g}|\leq\sqrt{x_{1}x_{2}}\sum\limits_{s=1}
^{2}\left\{  \sum\limits_{M,N=\pm}\left(  y_{1}^{M}y_{2}^{N}\right)
^{2}\left|  \left[  L_{s}(G^{g})\right]  _{MN}\right|  ^{2}+2\left|  \det
L_{s}(G^{g})\right|  y_{1}^{+}y_{2}^{+}y_{1}^{-}y_{2}^{-}\right\}  ^{1/2}\,.
\end{equation}
Here we use the values $M,N=\pm$ for matrix indices instead of $1,2$.

According to Eq. (\ref{x-k-def}) we have
\begin{equation}
\sqrt{x_{1}x_{2}}\sqrt{1-\xi^{2}}=\sqrt{x^{2}-\xi^{2}}\,.
\end{equation}
Therefore
\begin{equation}
|G^{g}|\leq\sqrt{x^{2}-\xi^{2}}\sum\limits_{s=1}^{2}\left\{  \sum
\limits_{M,N=\pm}\left(  y_{1}^{M}y_{2}^{N}\right)  ^{2}\left|  \left[
L_{s}(G^{g})\right]  _{MN}\right|  ^{2}+2\left|  \det L_{s}(G^{g})\right|
y_{1}^{+}y_{2}^{+}y_{1}^{-}y_{2}^{-}\right\}  ^{1/2}\,.
\label{G-bound-general-gluon-2}
\end{equation}

\section{Derivation of positivity bounds for helicity-flip GPDs}

\label{L-G-matrices}

In order to apply the general positivity bounds (\ref{G-bound-general-2}) and
(\ref{G-bound-general-gluon-2}) to some GPD $G$, we must know the corresponding
matrices $L_{s}(G)$. In Appendix \ref{Rxample-L-k-calculation-appendix} we
describe how $L_{s}(G)$ can be computed using Eq. (\ref{G-L-B}). Below we list
the results.

\subsection{Matrices $L_{s}(G)$ for quarks}

Using notation
\begin{equation}
\eta_{s}=\left\{
\begin{array}
[c]{cc}
+1, & s=1,\\
-1, & s=2,
\end{array}
\right.
\end{equation}
we can write
\begin{equation}
L_{s}\left(  H_{T}^{q}\right)  =\frac{\alpha}{4}\left(
\begin{array}
[c]{cc}
\eta_{s} & -\nu\xi\\
-\nu\xi & -\eta_{s}
\end{array}
\right)  \,,\, \label{L-H-T-q}
\end{equation}
\begin{equation}
L_{s}\left(  \tilde{H}_{T}^{q}\right)  =\frac{\nu^{2}}{4\alpha}\left(
\begin{array}
[c]{cc}
0 & 0\\
0 & \eta_{s}
\end{array}
\right)  \,, \label{L-H-T-tilde-q}
\end{equation}
\begin{equation}
L_{s}\left(  E_{T}^{q}\right)  =\frac{\alpha\nu}{4}\left(
\begin{array}
[c]{cc}
0 & -1+\xi\\
1+\xi & -2\eta_{s}\nu
\end{array}
\right)  \,, \label{L-E-T-q}
\end{equation}
\begin{equation}
L_{s}\left(  \tilde{E}_{T}^{q}\right)  =\frac{\alpha\nu}{4}\left(
\begin{array}
[c]{cc}
0 & 1-\xi\\
1+\xi & -2\eta_{s}\nu\xi
\end{array}
\right)  \,. \label{L-E-T-tilde-q}
\end{equation}

\subsection{Matrices $L_{s}(G)$ for gluons}

In Appendix \ref{Rxample-L-k-calculation-appendix} we show that matrices
$L_{s}(G)$ for gluon GPDs $H_{T}^{g},\tilde{H}_{T}^{g},E_{T}^{g},\tilde{E}
_{T}^{g}$ differ from the corresponding quark matrices by the factor of $\nu$:
\begin{equation}
L_{s}\left(  G^{g}\right)  =\nu L_{s}(G^{q})\quad(G^g=H_{T}^g,\tilde{H}_{T}^g
,E_{T}^g,\tilde{E}_{T}^g)\,. \label{L-gluon-L-quark}
\end{equation}
Therefore
\begin{equation}
L_{s}\left(  H_{T}^{g}\right)  =\frac{\alpha\nu}{4}\left(
\begin{array}
[c]{cc}
\eta_{s} & -\nu\xi\\
-\nu\xi & -\eta_{s}
\end{array}
\right)  \,, \label{L-H-T-g}
\end{equation}
\begin{equation}
L_{s}\left(  \tilde{H}_{T}^{g}\right)  =\frac{\nu^{3}}{4\alpha}\left(
\begin{array}
[c]{cc}
0 & 0\\
0 & \eta_{s}
\end{array}
\right)  \,,
\end{equation}
\begin{equation}
L_{s}\left(  E_{T}^{g}\right)  =\frac{\alpha\nu^{2}}{4}\left(
\begin{array}
[c]{cc}
0 & -1+\xi\\
1+\xi & -2\eta_{s}\nu
\end{array}
\right)  \,, \label{L-E-T-g}
\end{equation}
\begin{equation}
L_{s}\left(  \tilde{E}_{T}^{g}\right)  =\frac{\alpha\nu^{2}}{4}\left(
\begin{array}
[c]{cc}
0 & 1-\xi\\
1+\xi & -2\eta_{s}\nu\xi
\end{array}
\right)  \,.\,
\end{equation}
Note that in the positivity bound (\ref{G-bound-general-gluon-2}) we use the
values $\pm$ for the matrix indices of $L_{s}$. The above explicit matrix
expressions assume the interpretation ``$+=1$'' , ``$-=2$''.

\subsection{Example: derivation of inequalities for $H_{T}^{q}$}

\label{Example-H-q-T-section}

Let us show how the general inequality (\ref{G-bound-general-2}) can be used
for the derivation of the positivity bound for the GPD $H_{T}^{q}$. Inserting
the explicit expression (\ref{L-H-T-q}) for $L_{s}\left(  H_{T}^{q}\right)  $
into Eq. (\ref{G-bound-general-2}), we find
\[
|H_{T}^{q}|\leq\frac{\alpha}{4}\left\{  \left[  \left(  R_{1}^{1}R_{2}
^{1}+R_{1}^{2}R_{2}^{2}\right)  ^{2}+\nu^{2}\xi^{2}\left(  R_{1}^{1}R_{2}
^{2}+R_{1}^{2}R_{2}^{1}\right)  ^{2}\right]  ^{1/2}\right.
\]
\begin{equation}
\left.  +\left[  \left(  R_{1}^{3}R_{2}^{3}+R_{1}^{4}R_{2}^{4}\right)
^{2}+\nu^{2}\xi^{2}\left(  R_{1}^{3}R_{2}^{4}+R_{1}^{4}R_{2}^{3}\right)
^{2}\right]  ^{1/2}\right\}  \,. \label{H-T-derivation}
\end{equation}
According to Eqs. (\ref{Q-4l-Q-2l}) and (\ref{R-jk-def}) we have
\begin{equation}
R_{s}^{4}=R_{s}^{2}\,.
\end{equation}
Now we see that inequality (\ref{H-T-derivation}) coincides with the
inequality (\ref{H-T-example-3}) which was already discussed above. As was
explained in Sec. \ref{Examples-section}, two other positivity bounds
for $H_{T}^{q}$ (\ref{H-T-example-1b}) and (\ref{H-T-example-2}) can be
derived from the already proved inequality (\ref{H-T-example-3}).

\section{Results}

\label{Results-section}

\subsection{Hierarchy of positivity bounds}

In this paper we have chosen quark GPD $H_{T}^{q}$ in order to illustrate
the methods used for the derivation of positivity bounds for GPDs.
We have derived three inequalities:

1) ``strong'' inequality (\ref{H-T-example-3}) where the GPD is bounded by a
combination of all FPDs (transversity, longitudinally polarized and unpolarized);

2) ``intermediate'' positivity bound (\ref{H-T-example-2}) where the GPD is
bounded by a combination of unpolarized and longitudinally polarized FPDs
but the transversity FPD does not appear;

3) ``weak'' positivity bound (\ref{H-T-example-1b}) which involves only the
unpolarized FPD.

From the mathematical point of view we have a set of inequalities which can be
derived one from another, starting from the strong inequality
(\ref{H-T-example-3}). However, currently the use of the strong positivity
bound is limited by the absence of reliable data on the transversity FPD. On
the other hand, the intermediate and weak positivity bounds do not contain the
transversity FPD and can be used in practical applications.

The above classification of positivity bounds for GPDs was described in terms
of the quark GPD $H_{T}^{q}$ but the generalization for other quark GPDs is straightforward.

In the case of gluon GPDs the situation is simpler. Since there is
no gluon transversity FPD for the nucleon, we have only two types of
positivity bounds for gluon GPDs:

i) ``strong'' positivity bounds containing both polarized and unpolarized
gluon FPDs,

ii) ``weak'' positivity bounds involving only the unpolarized gluon FPD.

\subsection{Strong positivity bounds for GPDs}

\subsubsection{Strong positivity bounds for quark GPDs}

As was explained above, we use the word ``strong'' for those inequalities
which come directly from the general positivity bound (\ref{G-bound-general})
and contain all types of FPDs on the RHS. The set of ``strong''
inequalities for quark GPDs is
\begin{align}
|H_{T}^{q}|  &  \leq\frac{\alpha}{4}\sum\limits_{k=1,3}\left[  \left(
R_{1}^{k}R_{2}^{k}+R_{1}^{2}R_{2}^{2}\right)  ^{2}+\nu^{2}\xi^{2}\left(
R_{1}^{k}R_{2}^{2}+R_{1}^{2}R_{2}^{k}\right)  ^{2}\right]  ^{1/2}
\,,\label{H-T-q-strong}\\
|\tilde{H}_{T}^{q}|  &  \leq\frac{\nu^{2}}{2\alpha}R_{1}^{2}R_{2}
^{2}\,,\label{H-tilde-T-q-strong}\\
|E_{T}^{q}|  &  \leq\frac{\nu}{4\alpha}\sum\limits_{k=1,3}\left[  \left(
\frac{R_{1}^{k}R_{2}^{2}}{1+\xi}+\frac{R_{1}^{2}R_{2}^{k}}{1-\xi}\right)
^{2}+4\nu^{2}\alpha^{4}\left(  R_{1}^{2}R_{2}^{2}\right)  ^{2}\right]
^{1/2}\,,\label{E-T-q-strong}\\
|\tilde{E}_{T}^{q}|  &  \leq\frac{\nu}{4\alpha}\sum\limits_{k=1,3}\left[
\left(  \frac{R_{1}^{k}R_{2}^{2}}{1+\xi}+\frac{R_{1}^{2}R_{2}^{k}}{1-\xi
}\right)  ^{2}+4\nu^{2}\xi^{2}\alpha^{4}\left(  R_{1}^{2}R_{2}^{2}\right)
^{2}\right]  ^{1/2}\,. \label{E-tilde-T-q-strong}
\end{align}
Parameters $\alpha$ and $\nu$ are given by Eqs. (\ref{alpha-beta-def-0}) and
(\ref{gamma-def}).
The first inequality (\ref{H-T-q-strong}) is nothing else but Eq.
(\ref{H-T-example-3}) derived above. The derivation of the other inequalities is similar.
In all these inequalities, the GPDs on the LHS are taken at values $x,\xi,t$
[e.g. $H_{T}^{q}=H_{T}^{q}(x,\xi,t)$]. The FPDs appear in these bounds at
points $x_{k}$ (\ref{x-k-def}) via the quantities $R_{k}^{i}$. According to
Eqs. (\ref{R-jk-def}) and (\ref{Q1-4-def}) we have
\begin{equation}
\left(
\begin{array}
[c]{c}
R_{k}^{1}\\
R_{k}^{2}\\
R_{k}^{3}
\end{array}
\right)  =\left(
\begin{array}
[c]{c}
\sqrt{q(x_{k})+\Delta_{L}q(x_{k})+2\Delta_{T}q(x_{k})}\\
\sqrt{q(x_{k})-\Delta_{L}q(x_{k})}\\
\sqrt{q(x_{k})+\Delta_{L}q(x_{k})-2\Delta_{T}q(x_{k})}
\end{array}
\right)  \,. \label{R-k-again}
\end{equation}

\subsubsection{Strong positivity bounds for gluon GPDs}

\label{sec:gluon}

The difference between the quark and gluon cases is so small that the
inequalities for gluon GPDs can be directly obtained from quark inequalities
(\ref{H-T-q-strong})--(\ref{E-tilde-T-q-strong}) by a simple replacement. The
modification consists of two factors:

1) factor
\begin{equation}
w\equiv\sqrt{x^{2}-\xi^{2}}
\end{equation}
arising from the difference between the underlying general
quark (\ref{G-bound-general-2})
and gluon (\ref{G-bound-general-gluon-2}) inequalities,

2) factor $\nu$ coming from the relation (\ref{L-gluon-L-quark}) between the
quark and gluon matrices $L_{k}(G)$.

Taken together, these two factors give $\nu w$.
Now we must insert this factor into the quark inequalities
(\ref{H-T-q-strong})--(\ref{E-tilde-T-q-strong}) and replace
\begin{equation}
\left(
\begin{array}
[c]{c}
R_{k}^{1}\\
R_{k}^{2}\\
R_{k}^{3}
\end{array}
\right)  \rightarrow\left(
\begin{array}
[c]{c}
y_{k}^{+}\\
y_{k}^{-}\\
y_{k}^{+}
\end{array}
\right)  \,,
\end{equation}
where according to Eqs. (\ref{gpm}) and (\ref{s-k-pm-def}) we have
\begin{equation}
y_{k}^{\pm}=\sqrt{g(x_{k})\pm\Delta g(x_{k})}\,.
\end{equation}
As a result, we obtain
\begin{align}
|H_{T}^{g}|  &  \leq\frac{\nu\alpha w}{2}\left[  \left(  y_{1}^{+}y_{2}
^{+}+y_{1}^{-}y_{2}^{-}\right)  ^{2}+\nu^{2}\xi^{2}\left(  y_{1}^{-}y_{2}
^{+}+y_{1}^{+}y_{2}^{-}\right)  ^{2}\right]  ^{1/2}\,,\label{HgT1}\\
|\tilde{H}_{T}^{g}|  &  \leq\frac{\nu^{3}w}{2\alpha}y_{1}^{-}y_{2}^{-},\\
|E_{T}^{g}|  &  \leq\frac{\nu^{2}w}{2\alpha}\left[  \left(  \frac{y_{1}
^{+}y_{2}^{-}}{1+\xi}+\frac{y_{1}^{-}y_{2}^{+}}{1-\xi}\right)  ^{2}+4\nu
^{2}\alpha^{4}(y_{1}^{-}y_{2}^{-})^{2}\right]  ^{1/2}\,,\\
|\tilde{E}_{T}^{g}|  &  \leq\frac{\nu^{2}w}{2\alpha}\left[  \left(
\frac{y_{1}^{+}y_{2}^{-}}{1+\xi}+\frac{y_{1}^{-}y_{2}^{+}}{1-\xi}\right)
^{2}+4\nu^{2}\alpha^{4}\xi^{2}(y_{1}^{-}y_{2}^{-})^{2}\right]  ^{1/2}\,.
\end{align}

\subsection{Consequences from strong positivity bounds}

\subsubsection{Positivity bounds for quark GPDs without forward transversity distribution}

The elimination of the transversity FPD from the positivity bounds for quark
GPDs is discussed in Appendix \ref{Transversity-exclusion-appendix}. Here we
just list the results:
\begin{align}
|H_{T}^{q}|  &  \leq\frac{\alpha}{2}\left[  \left(  
r_{1}^{+}r_{2}^{+}
+r_{1}^{-}r_{2}^{-}\right)^{2}
+\nu^{2}\xi^{2}\left(  r_{1}^{-}r_{2}^{+}
+r_{1}^{+}r_{2}^{-}\right)  ^{2}\right]  ^{1/2}\,,
\label{H-T-intermediate}\\
|\tilde{H}_{T}^{q}|  &  \leq\frac{\nu^{2}}{2\alpha}r_{1}^{-}r_{2}
^{-}\,,\label{Hq-T-tilde-intermediate}\\
|E_{T}^{q}|  &  \leq\frac{\nu}{2\alpha}
\left[
\left(  \frac{r_{1}^{+}r_{2}^{-}}{1+\xi}
+\frac{r_{1}^{-}r_{2}^{+}}{1-\xi}\right)  ^{2}+4\nu^{2}
\alpha^{4}\left(  r_{1}^{-}r_{2}^{-}\right)  ^{2}
\right]^{1/2}
\,,\label{E-T-q-org}\\
|\tilde{E}_{T}^{q}|  &  \leq\frac{\nu}{2\alpha}
\left[
\left(
\frac{r_{1}^{+}r_{2}^{-}}{1+\xi}
+\frac{r_{1}^{-}r_{2}^{+}}{1-\xi}\right)
^{2}+4\nu^{2}\alpha^{4}\xi^{2}\left(  r_{1}^{-}r_{2}^{-}\right)  ^{2}
\right]^{1/2}\,.
\label{E-T-tilde-q-org}
\end{align}
Parameters $r_{k}^{\pm}$ appearing on the RHS were defined in Eq.
(\ref{r-k-pm-def}). They depend on FPDs $q$ and $\Delta_{L}q$ only.

\subsubsection{Positivity bounds for GPDs containing only unpolarized FPDs}

The positivity bound for $H_{T}^{q}$ (\ref{H-T-example-1b})
containing only the unpolarized FPD has
been already derived from Eq. (\ref{H-T-intermediate}) in Section
\ref{Enhanced-HTq-section}. A similar inequality for $\tilde{H}_{T}^{q}$ can
be obtained from inequality Eq. (\ref{Hq-T-tilde-intermediate}) by noticing
that $r_{k}^{\pm}\leq\sqrt{2q_{k}}$. Thus, for the quark GPDs we have
\begin{align}
|H_{T}^{q}|  &  \leq\alpha\sqrt{(1+\nu^{2}\xi^{2})q_{1}q_{2}}\,,\\
|\tilde{H}_{T}^{q}|  &  \leq\nu^{2}\alpha^{-1}\sqrt{q_{1}q_{2}}\,
\end{align}
with the $q_{k}$ defined in Eq. (\ref{q-k-def}).

For gluons we similarly find
\begin{align}
|H_{T}^{g}|  &  \leq\nu w\alpha\sqrt{(1+\nu^{2}\xi^{2})g_{1}g_{2}
},\label{HgTshort}\\
|\tilde{H}_{T}^{g}|  &  \leq\nu^{3}\alpha^{-1}w\sqrt{g_{1}g_{2}},
\label{HgTtshort}
\end{align}
where $g_{k}=g(x_{k})$.

\section{Other bounds and application to DVCS}

Helicity-flip gluon GPDs appear in the QCD description of DVCS (via the
interference of the DVCS amplitude with the Bethe-Heitler process)
\cite{Diehl:2001pm,BMK-2001}. Helicity-flip quark GPDs can also be accessed in the
diffractive electroproduction of two vector mesons
\cite{IPST-02,IPST-03,IPST-04}. In practical applications one usually deals
with linear combinations of GPDs. In principle, we can simply combine the
trivial inequality
\begin{equation}
\left|  c_{1}H_{T}+c_{2}\tilde{H}_{T}+c_{3}E_{T}+c_{4}\tilde{E}_{T}\right|
\leq\left|  c_{1}H_{T}\right|  +\left|  c_{2}\tilde{H}_{T}\right|  +\left|
c_{3}E_{T}\right|  +\left|  c_{4}\tilde{E}_{T}\right|
\end{equation}
with the above bounds for separate GPDs $H_{T},E_{T},\tilde{H}_{T},\tilde
{E}_{T}$. However, one can derive a \emph{stronger} bound for the same linear
combination using the basic inequalities (\ref{G-bound-general-2}),
(\ref{G-bound-general-gluon-2}).

In order to illustrate this method, let us consider the interference of the
Bethe-Heitler and DVCS amplitudes in the case of unpolarized protons and
electrons. The $\cos3\phi$ interference term can be expressed via the
following linear combination of gluon helicity-flip GPDs \cite{Diehl:2001pm,BMK-2001}
\begin{equation}
G_{3\phi}^{\mathrm{DVCS}}=F_{2}H_{T}^{g}-F_{1}E_{T}^{g}-2\left(
F_{1}+\frac{t}{4m^{2}}F_{2}\right)  \tilde{H}_{T}^{g}\,.
\end{equation}
Here $F_{1,2}$ are Dirac form factors.
In order to apply the general positivity bound (\ref{G-bound-general-gluon-2})
to $G_{3\phi}^{\mathrm{DVCS}}$, we first must compute the corresponding matrix
$L_{s}(G_{3\phi}^{\mathrm{DVCS}})$:
\begin{equation}
L_{s}(G_{3\phi}^{\mathrm{DVCS}})=F_{2}L_{s}\left(  H_{T}^{g}\right)
-F_{1}L_{s}\left(  E_{T}^{g}\right)  -2\left(  F_{1}+\frac{t}{4m^{2}}
F_{2}\right)  L_{s}\left(  \tilde{H}_{T}^{g}\right)  \,.
\end{equation}
Inserting Eqs. (\ref{L-H-T-g})--(\ref{L-E-T-g}) and using expression
(\ref{t-nu-xi}) for $t$, we find
\begin{equation}
L_{s}(G_{3\phi}^{\mathrm{DVCS}})=\frac{\alpha\nu}{4}\left(
\begin{array}
[c]{ll}
\eta_{s}F_{2} & \nu\left[  -\xi F_{2}+\left(  1-\xi\right)  F_{1}\right] \\
-\nu\left[  \xi F_{2}+\left(  1+\xi\right)  F_{1}\right]  \quad & \eta
_{s}\left[  (1+2\nu^{2}\xi^{2})F_{2}+2F_{1}\nu^{2}\xi^{2}\right]
\end{array}
\right)  .
\end{equation}
Let us apply the general formula (\ref{G-bound-general-gluon-2}):
\begin{equation}
|G_{3\phi}^{\mathrm{DVCS}}|\leq w\sum\limits_{s=1}^{2}\left\{  \sum
\limits_{M,N=\pm}\left(  y_{1}^{M}y_{2}^{N}\right)  ^{2}\left|  \left[
L_{s}\left(  G_{3\phi}^{\mathrm{DVCS}}\right)  \right]  _{MN}\right|
^{2}+2\left|  \det L_{s}\left(  G_{3\phi}^{\mathrm{DVCS}}\right)  \right|
y_{1}^{+}y_{2}^{+}y_{1}^{-}y_{2}^{-}\right\}  ^{1/2}\,.
\label{aux-3-phi}
\end{equation}
Using notation
\begin{equation}
g^{\pm}_k=g^{\pm}(x_k)=(y^{\pm}_k)^2
\,,
\end{equation}
we obtain
\begin{align}
&  \sum\limits_{M,N=\pm}\left(  y_{1}^{M}y_{2}^{N}\right)  ^{2}\left|
\left[  L_{s}(G_{3\phi}^{\mathrm{DVCS}})\right]  _{MN}\right|  ^{2}\nonumber\\
&  =\left.  \left(  \frac{\alpha\nu}{4}\right)  ^{2}\right\{  \left(
F_{2}\right)  ^{2}g_{1}^{+}g_{2}^{+}+\left[  -\nu\xi F_{2}+\left(
1-\xi\right)  F_{1}\nu\right]  ^{2}g_{1}^{+}g_{2}^{-}+\left[  -\nu\xi
F_{2}-\left(  1+\xi\right)  F_{1}\nu\right]  ^{2}g_{2}^{+}g_{1}^{-}\nonumber\\
&  \left.  +\left[  2F_{1}\nu^{2}\xi^{2}+F_{2}\left(  1+2\nu^{2}\xi
^{2}\right)  \right]  ^{2}g_{1}^{-}g_{2}^{-}\right\}  \,,
\end{align}
\begin{equation}
\det L_{s}(G_{3\phi}^{\mathrm{DVCS}})=\left(  \frac{\alpha\nu}{4}\right)
^{2}\left[  \left(  F_{1}\nu\alpha^{-1}\right)  ^{2}+\left(  F_{2}\right)
^{2}\left(  1+\nu^{2}\xi^{2}\right)  \right]  \,.
\end{equation}
As a result, we find from Eq. (\ref{aux-3-phi})
\begin{align}
\left|  G_{3\phi}^{\mathrm{DVCS}}\right|   &  \leq\left.  \frac{w\alpha\nu}
{2}\right\{  \left(  F_{2}\right)  ^{2}g_{1}^{+}g_{2}^{+}\nonumber\\
&  +\nu^{2}\left[  \xi F_{2}-\left(  1-\xi\right)  F_{1}\right]  ^{2}g_{1}
^{+}g_{2}^{-}+\nu^{2}\left[  \xi F_{2}+\left(  1+\xi\right)  F_{1}\right]
^{2}g_{2}^{+}g_{1}^{-}\nonumber\\
&  +\left[  2F_{1}\nu^{2}\xi^{2}+F_{2}\left(  1+2\nu^{2}\xi^{2}\right)
\right]  ^{2}g_{1}^{-}g_{2}^{-}\nonumber\\
&  \left.  +2\left[  \left(  F_{1}\nu\alpha^{-1}\right)  ^{2}+\left(
F_{2}\right)  ^{2}\left(  1+\nu^{2}\xi^{2}\right)  \right]  \sqrt{g_{1}
^{+}g_{2}^{+}g_{1}^{-}g_{2}^{-}}\right\}  ^{1/2}\,.
\end{align}

\section{Conclusions}

In this paper we have derived a set of positivity bounds for the quark and
gluon helicity-flip GPDs. For every GPD we have several positivity bounds
ranging from the simplest bounds in terms of unpolarized forward distributions
to the stronger bounds involving the transversity forward distribution.

It should be stressed that the method developed in
this paper can be more useful than the bounds presented here. Indeed, in
practical applications one usually deals with linear combinations of GPDs
depending on the specific process. The method described in this paper allows
one to derive positivity bounds for any linear combination of GPDs.

The power of positivity bounds is limited to the region $|x|>\xi$. Although we
have no explicit positivity bounds at $|x|<\xi$, this region is not completely
isolated and free of any constraints. Indeed, the polynomiality of GPDs
imposes restrictions involving both regions. It should be also emphasized that
the inequalities derived in this paper are just a small part of the general positivity
bounds which can be written in terms of the impact parameter representation
\cite{Pobylitsa-02-c}.

Thus, one has an infinite amount of the polynomiality and positivity
constraints (in the impact parameter representation). The construction of the
complete general representation for GPDs obeying all these constraints remains
a challenging problem.

\acknowledgments    

We appreciate discussions with A.V. Belitsky, V. Braun, J.C. Collins, M.
Diehl, A.V. Efremov, L. Frankfurt, D. Kiptily, N. Kivel, D. M{\"u}ller, M.V.
Polyakov, A.V. Radyushkin, M. Strikman, and O. Teryaev. This work was
supported by DFG and BMBF. M.K. acknowledges the support of the GRK-814 (DFG).

\appendix                                

\section{Helicity amplitudes}

\subsection{Definition}

\label{Helicity-amplitudes-appendix}

It is convenient to analyze the general positivity bounds for GPDs using
``helicity amplitudes'' introduced in Ref.~\cite{Diehl:2001pm}
\begin{equation}
A_{\lambda^{\prime}\mu^{\prime},\lambda\mu}=\left.  \int\frac{dz^{-}}{2\pi
}e^{ixP^{+}z^{-}}\langle p^{\prime},\lambda^{\prime}|O_{\mu^{\prime},\mu
}(z)|p,\lambda\rangle\right|  _{z^{+}=0,z^{\perp}=0}\,. \label{A-amplitudes}
\end{equation}
Here $O_{\mu^{\prime},\mu}$ are bilinear quark light-ray operators with the
polarization indices $\mu,\mu^{\prime}$. Next, $|p,\lambda\rangle$ is a
nucleon state with momentum $p$ and polarization $\lambda$ (in the sense of
light-cone helicity states \cite{KS-70}), $P^{+}$ is the light-cone component
of vector $P=(p+p^{\prime})/2$.

With two values for each polarization index of $A_{\lambda^{\prime}\mu
^{\prime},\lambda\mu}$ we have $2^{4}=16$ components but due to the parity
invariance \cite{Diehl:2001pm}
\begin{equation}
A_{-\lambda^{\prime},-\mu^{\prime};-\lambda,-\mu}=(-1)^{\lambda^{\prime}
-\mu^{\prime}-\lambda+\mu}A_{\lambda^{\prime}\mu^{\prime},\lambda\mu}
\label{helicity-reflection}
\end{equation}
only eight components are independent.
This is exactly the number of twist-two GPDs (\ref{twist-2-GPDs}).

\subsection{Quark helicity amplitudes}

We take the expressions for the helicity amplitudes $A_{\lambda^{\prime
}\mu^{\prime},\lambda\mu}^{q}$ via GPDs from Ref. \cite{Diehl:2001pm} using
variables $\alpha$ (\ref{alpha-beta-def-0})
and $\nu$ (\ref{gamma-def}):
\begin{align}
A_{++,++}^{q}  &  =\frac{1}{2\alpha}\left[  \left(  H^{q}+\tilde{H}
^{q}\right)  -\alpha^{2}\xi^{2}\left(  E^{q}+\tilde{E}^{q}\right)  \right]
\,,\label{A-pp-pp}\\
A_{-+,-+}^{q}  &  =\frac{1}{2\alpha}\left[  \left(  H^{q}-\tilde{H}
^{q}\right)  -\alpha^{2}\xi^{2}\left(  E^{q}-\tilde{E}^{q}\right)  \right]
\,,\label{A-mp-mp}\\
A_{++,-+}^{q}  &  =-\frac{\alpha}{2\nu}\left(  E^{q}-\xi\tilde{E}^{q}\right)
\,,\label{A-pp-mp}\\
A_{-+,++}^{q}  &  =\frac{\alpha}{2\nu}\left(  E^{q}+\xi\tilde{E}^{q}\right)
\,, \label{A-mp-pp}
\end{align}
\begin{align}
A_{++,+-}^{q}  &  =\alpha\nu^{-1}\left[  \tilde{H}_{T}^{q}+\frac{1-\xi}
{2}\left(  E_{T}^{q}+\tilde{E}_{T}^{q}\right)  \right]  \,,\label{A-pp-pm}\\
A_{-+,--}^{q}  &  =\alpha\nu^{-1}\left[  \tilde{H}_{T}^{q}+\frac{1+\xi}
{2}\left(  E_{T}^{q}-\tilde{E}_{T}^{q}\right)  \right]  \,,\label{A-mp-mm}\\
A_{++,--}^{q}  &  =\alpha\left(  \alpha^{-2}H_{T}^{q}+\nu^{-2}\tilde{H}
_{T}^{q}-\xi^{2}E_{T}^{q}+\xi\tilde{E}_{T}^{q}\right)  \,,\label{A-pp-mm}\\
A_{-+,+-}^{q}  &  =-\alpha\nu^{-2}\tilde{H}_{T}^{q}\,. \label{A-mp-pm}
\end{align}

\subsection{Gluon helicity amplitudes}

Now we list the expressions for the gluon helicity amplitudes $A_{\lambda
^{\prime}\mu^{\prime},\lambda\mu}^{g}$ via GPDs from Ref. \cite{Diehl:2001pm}
(taking $\varepsilon=+1$ in equations of Ref. \cite{Diehl:2001pm}). Amplitudes which conserve
the gluon helicity have the same form as in the quark case:
\begin{align}
A_{++,++}^{g}  &  =\frac{1}{2\alpha}\left[  \left(  H^{g}+\tilde{H}
^{g}\right)  -\alpha^{2}\xi^{2}\left(  E^{g}+\tilde{E}^{g}\right)  \right]
\,,\label{A-g-pppp}\\
A_{-+,-+}^{g}  &  =\frac{1}{2\alpha}\left[  \left(  H^{g}-\tilde{H}
^{g}\right)  -\alpha^{2}\xi^{2}\left(  E^{g}-\tilde{E}^{g}\right)  \right]
\,,\\
A_{++,-+}^{g}  &  =-\frac{\alpha}{2\nu}\left(  E^{g}-\xi\tilde{E}^{g}\right)
\,,\\
A_{-+,++}^{g}  &  =\frac{\alpha}{2\nu}\left(  E^{g}+\xi\tilde{E}^{g}\right)
\,, \label{A-g-mppp}
\end{align}
but the helicity-flip amplitudes differ by a factor of $\nu^{-1}$
\begin{align}
A_{++,+-}^{g}  &  =\alpha\nu^{-2}\left[  \tilde{H}_{T}^{g}+\frac{1-\xi}
{2}\,\left(  E_{T}^{g}+\tilde{E}_{T}^{g}\right)  \right]  ,\label{A-g-pppm}\\
A_{-+,--}^{g}  &  =\alpha\nu^{-2}\left[  \tilde{H}_{T}^{g}+\frac{1+\xi}
{2}\,\left(  E_{T}^{g}-\tilde{E}_{T}^{g}\right)  \right]  ,\\
A_{++,--}^{g}  &  =\alpha\nu^{-1}\left(  \alpha^{-2}H_{T}^{g}+\ \nu^{-2}
\tilde{H}_{T}^{g}-\xi^{2}\,E_{T}^{g}+\xi\,\tilde{E}_{T}^{g}\right)  ,\\
A_{-+,+-}^{g}  &  =-\,\alpha\nu^{-3}\tilde{H}_{T}^{g}\,. \label{A-g-mppm}
\end{align}

\subsection{Expressions for quark GPDs via helicity amplitudes}

Solving the system of Eqs. (\ref{A-pp-pm})--(\ref{A-mp-pm}) with respect
to GPDs, we find
\begin{align}
H_{T}^{q}  &  =\alpha\left(  A_{++,--}^{q}+\ A_{-+,+-}^{q}\right)  +\alpha
\nu\xi\left(  A_{-+,--}^{q}-A_{++,+-}^{q}\right)  \,,\label{H-q-T-via-A}\\
\tilde{H}_{T}^{q}  &  =-\alpha^{-1}\nu^{2}A_{-+,+-}^{q}
\,,\label{H-q-T-tilde-via-A}\\
E_{T}^{q}  &  =\alpha\nu\left[  (1-\xi)A_{-+,--}^{q}+(1+\xi)A_{++,+-}
^{q}\right]  +2\alpha\nu^{2}A_{-+,+-}^{q}\,,\\
\tilde{E}_{T}^{q}  &  =\alpha\nu\left[  (1+\xi)A_{++,+-}^{q}-(1-\xi
)A_{-+,--}^{q}\right]  +2\alpha\xi\nu^{2}A_{-+,+-}^{q}\,.
\label{E-q-tilde-T-via-A}
\end{align}

\subsection{Expressions for gluon GPDs via helicity amplitudes}

Now we want to solve equations (\ref{A-g-pppm})--(\ref{A-g-mppm}) with respect to
the gluon GPDs. Note that the only difference form the quark equations is the extra
factor of $\nu^{-1}$ in Eqs. (\ref{A-g-pppm})--(\ref{A-g-mppm}). Therefore we
can read the gluon results form quark expressions
(\ref{H-q-T-via-A})--(\ref{E-q-tilde-T-via-A}):
\begin{align}
H_{T}^{g}  &  =\alpha\nu\left(  A_{++,--}^{g}+\ A_{-+,+-}^{g}\right)
+\alpha\nu^{2}\xi\left(  A_{-+,--}^{g}-A_{++,+-}^{g}\right)
\,,\label{H-g-T-via-A}\\
\tilde{H}_{T}^{g}  &  =-\alpha^{-1}\nu^{3}A_{-+,+-}^{g}\,,\\
E_{T}^{g}  &  =\alpha\nu^{2}\left[  (1-\xi)A_{-+,--}^{g}+(1+\xi)A_{++,+-}
^{g}\right]  +2\alpha\nu^{3}A_{-+,+-}^{g}\,,\\
\tilde{E}_{T}^{g}  &  =\alpha\nu^{2}\left[  (1+\xi)A_{++,+-}^{g}
-(1-\xi)A_{-+,--}^{g}\right]  +2\alpha\xi\nu^{3}A_{-+,+-}^{g}\,.
\label{E-g-tilde-T-via-A}
\end{align}

\subsection{From $A$ to $B_{k}$}
\label{A-B-appendix}

Instead of the tensor $A_{\lambda^{\prime}\mu^{\prime},\lambda\mu}$, it is
convenient to use the matrix $\tilde{A}_{ab}$ defined by the linear
transformation
\begin{equation}
\tilde{A}_{ab}=\sum\limits_{\lambda^{\prime}\mu^{\prime}\lambda\mu}\left(
V_{a}^{\lambda^{\prime}\mu^{\prime}}\right)  ^{\ast}A_{\lambda^{\prime}
\mu^{\prime},\lambda\mu}V_{b}^{\lambda\mu}\quad(a,b=1,2,3,4)\,,
\label{A-tilde-via-A}
\end{equation}
where $V_{b}^{\lambda\mu}$ has following nonzero components
\begin{equation}
V_{1}^{++}=V_{3}^{++}=V_{1}^{--}=-V_{3}^{--}=V_{2}^{+-}=V_{4}^{+-}=V_{4}
^{-+}=-V_{2}^{-+}=1\,. \label{O-matrix-def}
\end{equation}
Matrix $\tilde{A}_{ab}$ has a block diagonal structure
\begin{equation}
\tilde{A}=\left(
\begin{array}
[c]{cc}
B_{1} & 0\\
0 & B_{2}
\end{array}
\right)  \,, \label{A-tilde-via-B}
\end{equation}
where the $2\times2$ matrices $B_{1}$, $B_{2}$ are
\begin{equation}
B_{1}=\left(
\begin{array}
[c]{cc}
\tilde{A}_{11} & \tilde{A}_{12}\\
\tilde{A}_{21} & \tilde{A}_{22}
\end{array}
\right)  \,,\quad B_{2}=\left(
\begin{array}
[c]{cc}
\tilde{A}_{33} & \tilde{A}_{34}\\
\tilde{A}_{43} & \tilde{A}_{44}
\end{array}
\right)  \,. \label{B-Atilde}
\end{equation}
Now we find from Eqs. (\ref{helicity-reflection}), (\ref{A-tilde-via-A}),
(\ref{O-matrix-def}) and (\ref{B-Atilde})
\begin{align}
A_{++,--}  &  =\frac{1}{4}\left[  (B_{1})_{11}-(B_{2})_{11}\right]
\,,\label{A-pp-mm-via-B}\\
A_{-+,--}  &  =-\frac{1}{4}\left[  (B_{1})_{21}+(B_{2})_{21}\right]  \,,\\
A_{++,+-}  &  =\frac{1}{4}\left[  (B_{1})_{12}+(B_{2})_{12}\right]  \,,\\
A_{-+,+-}  &  =\frac{1}{4}\left[  (B_{2})_{22}-(B_{1})_{22}\right]  \,.
\label{A-mp-pm-via-B}
\end{align}

\subsection{Example of a calculation of $L_{k}\left(  G\right)  $}

\label{Rxample-L-k-calculation-appendix}

Let us illustrate the calculation of matrices $L_{k}\left(  G\right)  $ for
the GPD $H_{T}^{q}$. Inserting
Eqs. (\ref{A-pp-mm-via-B})--(\ref{A-mp-pm-via-B}) into expression (\ref{H-q-T-via-A}) for $H_{T}^{q}$,
we obtain
\begin{align}
H_{T}^{q}  &  =\frac{\alpha}{4}\left\{  \left[  (B_{1})_{11}-(B_{2}
)_{11}\right]  +\left[  (B_{2})_{22}-(B_{1})_{22}\right]  \right\} \nonumber\\
&  -\frac{1}{4}\alpha\nu\xi\left\{  \left[  (B_{1})_{21}+(B_{2})_{21}\right]
+\left[  (B_{1})_{12}+(B_{2})_{12}\right]  \right\} \nonumber\\
&  =\frac{\alpha}{4}\mathrm{Tr}\left[  \left(
\begin{array}
[c]{cc}
1 & -\nu\xi\\
-\nu\xi & -1
\end{array}
\right)  B_{1}+\left(
\begin{array}
[c]{cc}
-1 & -\nu\xi\\
-\nu\xi & 1
\end{array}
\right)  B_{2}\right]  \,.
\end{align}
Comparing this result with the general representation (\ref{G-L-B}), we arrive
at the result (\ref{L-H-T-q}) for $L_{k}\left(  H_{T}^{q}\right)$. The
other quark matrices $L_{k}\left(G^q\right)  $
(\ref{L-H-T-tilde-q})--(\ref{L-E-T-tilde-q}) can be computed in the same way using equations
(\ref{H-q-T-tilde-via-A})--(\ref{E-q-tilde-T-via-A}) for $\tilde{H}_{T}^{q}$,
$E_{T}^{q}$, $\tilde{E}_{T}^{q}$.

The gluon matrices $L_{k}(G^{g})$ can be immediately read from the quark
matrices $L_{k}(G^{q})$, using relation (\ref{L-gluon-L-quark}). This relation
follows from the fact that expressions
(\ref{H-g-T-via-A})--(\ref{E-g-tilde-T-via-A}) for gluon GPDs differ from the corresponding
quark expressions (\ref{H-q-T-via-A})--(\ref{E-q-tilde-T-via-A}) only by a
factor of $\nu^{-1}$.

\section{Exclusion of the forward transversity distribution from the
positivity bounds}

\label{Transversity-exclusion-appendix}

\subsection{Bound on $H_{T}^{q}$}

In section \ref{H-stong-bound-example-section} it was announced that
inequality (\ref{H-T-example-2}) can be derived from the stronger inequality
(\ref{H-T-example-3}). In this appendix we will demonstrate this.

Introducing the variables
\begin{equation}
c_{jk}=\left(  \frac{R_{k}^{j}}{R_{k}^{2}}\right)  ^{2} \label{cd-2}
\end{equation}
and the function
\begin{equation}
h(v_{1},v_{2})=\left[  \left(  1+\sqrt{v_{1}v_{2}}\right)  ^{2}+\nu^{2}\xi
^{2}\left(  \sqrt{v_{1}}+\sqrt{v_{2}}\right)  ^{2}\right]  ^{1/2}\,,
\label{g-vv-def}
\end{equation}
we can rewrite inequality (\ref{H-T-example-3}) in the form
\begin{equation}
|H_{T}^{q}|\leq\frac{\alpha}{4}R_{1}^{2}R_{2}^{2}\left[  h(c_{11}
,c_{12})+h(c_{31},c_{32})\right]  \,. \label{H-via-g}
\end{equation}
Obviously $c_{jk}\ge 0$ here. Now we want to show that function $h$ is convex:
\begin{equation}
\left\|  -\frac{\partial^{2}h(v_{1},v_{2})}{\partial v_{k}\partial v_{l}
}\right\|  \geq0\,. \label{d2-h-positive}
\end{equation}

Let us define
\begin{equation}
\phi(v_{1},v_{2})=\left(  1+\sqrt{v_{1}v_{2}}\right)  ^{2}+\left(  \nu
\xi\right)  ^{2}\left(  \sqrt{v_{1}}+\sqrt{v_{2}}\right)  ^{2}
\end{equation}
so that
\begin{equation}
h(v_{1},v_{2})=\sqrt{\phi(v_{1},v_{2})}\,.
\end{equation}
Then
\begin{equation}
-\frac{\partial^2 h(v_{1},v_{2})}{\partial v_{k}\partial v_{l}}=\frac{1}
{4\phi^{3/2}}\left[  \frac{\partial\phi}{\partial v_{k}}\frac{\partial\phi
}{\partial v_{l}}-2\phi\frac{\partial^2\phi}{\partial v_{k}\partial v_{l}
}\right]  \,. \label{d2-h-d2-phi}
\end{equation}
Thus, in order to prove the convexity of $h$, we must check
that matrix
\begin{equation}
D_{kl}=\frac{\partial\phi}{\partial v_{k}}\frac{\partial\phi}{\partial v_{l}
}-2\phi\frac{\partial^2\phi}{\partial v_{k}\partial v_{l}}
\end{equation}
is positive definite:
\begin{equation}
\left\|  D_{kl}\right\|  \geq0\,.
\end{equation}
The positivity of the first term is obvious:
\begin{equation}
\left\|  \frac{\partial\phi}{\partial v_{k}}\frac{\partial\phi}{\partial
v_{l}}\right\|  \geq0\,.
\end{equation}
Now let us show that the second term
\begin{equation}
\left\|  -2\phi\frac{\partial^{2}\phi}{\partial v_{k}\partial v_{l}}\right\|
\geq0
\end{equation}
is also positive definite. Since
\begin{equation}
\phi\geq0\,,
\end{equation}
it is enough to show that
\begin{equation}
\left\|  -\frac{\partial^{2}\phi}{\partial v_{k}\partial v_{l}}\right\|
\geq0\,.
\end{equation}

Indeed,
\begin{equation}
\phi(v_{1},v_{2})=1+2\left[  1+\left(  \nu\xi\right)  ^{2}\right]  \sqrt
{v_{1}v_{2}}+\left(  \nu\xi\right)  ^{2}\left(  v_{1}+v_{2}\right)
\end{equation}
so that
\begin{equation}
\left\|  -\frac{\partial^{2}\phi}{\partial v_{k}\partial v_{l}}\right\|
=\frac{1}{2}\left[  1+\left(  \nu\xi\right)  ^{2}\right]
  \frac{1}{\sqrt{v_{1}v_{2}}}\left(
\begin{array}
[c]{cc}
\frac{v_{2}}{v_{1}} & -1\\
-1 & \frac{v_{1}}{v_{2}}
\end{array}
\right)  \,.
\end{equation}
Obviously this matrix is degenerate and non-negative. Combining this positivity
with Eq. (\ref{d2-h-d2-phi}), we complete the derivation of Eq.
(\ref{d2-h-positive}). This result means that function $h$
is convex. Therefore
\begin{equation}
h(c_{11},c_{12})+h(c_{31},c_{32})  \leq2h\left(  \frac{c_{11}
+c_{31}}{2},\frac{c_{12}+c_{32}}{2}\right)  \,.
\end{equation}
Together with inequality (\ref{H-via-g}) this yields
\begin{equation}
|H_{T}^{q}|\leq\frac{\alpha}{4}R_{1}^{2}R_{2}^{2}\left[  h(c_{11}
,c_{12})+h(c_{31},c_{32})\right]  \leq\frac{\alpha}{2}R_{1}^{2}R_{2}
^{2}h\left(  \frac{c_{11}+c_{31}}{2},\frac{c_{12}+c_{32}}{2}\right)  \,.
\label{H-qT-bound-4}
\end{equation}

According to Eqs. (\ref{r-k-pm-def}), (\ref{cd-2}), and (\ref{Q1-4-def})
we have
\begin{equation}
\frac{c_{1k}+c_{3k}}{2}=\frac12\left[
\left(  \frac{R_{k}^{1}}{R_{k}^{2}}\right)
^{2}+\left(  \frac{R_{k}^{3}}{R_{k}^{2}}\right)^{2}
\right]
=\frac{\left(
q+\Delta_{L}q\right)  _{k}}{\left(  q-\Delta_{L}q\right)  _{k}}=\left(
\frac{r_{k}^{+}}{r_{k}^{-}}\right)  ^{2}\,.
\end{equation}
Therefore inequality (\ref{H-qT-bound-4}) takes the form
\begin{equation}
|H_{T}^{q}|\leq\frac{\alpha}{2}r_{1}^{-}r_{2}^{-}h\left(  \left(
\frac{r_{1}^{+}}{r_{1}^{-}}\right)  ^{2},\left(  \frac{r_{2}^{+}}{r_{2}^{-}
}\right)  ^{2}\right)  \,.
\end{equation}
According to Eq. (\ref{g-vv-def}) this leads to inequality
(\ref{H-T-example-2}).

\subsection{Intermediate bounds on $E_{T}^{q}$ and $\tilde{E}_{T}^{q}$}

Now we want to repeat the work on the elimination of $\Delta_{T}q$ for other
quark positivity bounds. Note that the positivity bound
(\ref{H-tilde-T-q-strong}) for $\tilde{H}_{T}^{q}$ does not contain the
transversity distribution. Indeed, the RHS of this inequality depends on FPDs only
via $R_{k}^{2}$ and according to Eq. (\ref{R-k-again}) $R_{k}^{2}$ is
independent of $\Delta_{T}q$. Thus, the problem of the elimination of the
transversity must be solved only for GPDs $E_{T}^{q}$ and $\tilde{E}_{T}^{q}$.

We can rewrite inequality (\ref{E-T-q-strong}) in the form
\begin{equation}
|E_{T}^{q}|\leq\frac{1}{2}\left[  \sqrt{\left(  a_{1}\right)  ^{2}+b^{2}
}+\sqrt{\left(  a_{3}\right)  ^{2}+b^{2}}\right]  \,, \label{E-q-T-ab}
\end{equation}
where
\begin{equation}
a_{k}=\frac{\nu}{2\alpha}\left(  \frac{R_{1}^{k}R_{2}^{2}}{1+\xi}
+\frac{R_{1}^{2}R_{2}^{k}}{1-\xi}\right)  \,,
\end{equation}
\begin{equation}
b=\nu^{2}\alpha R_{1}^{2}R_{2}^{2}\,. \label{b-beta-def}
\end{equation}
Using the general inequality
\begin{equation}
\frac{1}{2}\left[  \sqrt{\left(  a_{1}\right)  ^{2}+b^{2}}+\sqrt{\left(
a_{3}\right)  ^{2}+b^{2}}\right]  \leq\sqrt{\frac{\left(  a_{1}\right)
^{2}+\left(  a_{3}\right)  ^{2}}{2}+b^{2}}\,,
\end{equation}
we find from inequality (\ref{E-q-T-ab})
\begin{equation}
|E_{T}^{q}|\leq\sqrt{\frac{\left(  a_{1}\right)  ^{2}+\left(  a_{3}\right)
^{2}}{2}+b^{2}}\,. \label{E-T-ineq-step-1}
\end{equation}
Here
\[
\left(  a_{1}\right)  ^{2}+\left(  a_{3}\right)  ^{2}=\frac{\nu^{2}}
{4\alpha^{2}}\left[  \left(  \frac{R_{1}^{1}R_{2}^{2}}{1+\xi}+\frac{R_{1}
^{2}R_{2}^{1}}{1-\xi}\right)  ^{2}+\left(  \frac{R_{1}^{3}R_{2}^{2}}{1+\xi
}+\frac{R_{1}^{2}R_{2}^{3}}{1-\xi}\right)  ^{2}\right]
\]
\begin{equation}
=\frac{\nu^{2}}{4\alpha^{2}}\left\{  \frac{\left[  \left(  R_{1}^{1}\right)
^{2}+\left(  R_{1}^{3}\right)  ^{2}\right]  \left(  R_{2}^{2}\right)  ^{2}
}{(1+\xi)^{2}}+\frac{\left[  \left(  R_{2}^{1}\right)  ^{2}+\left(  R_{2}
^{3}\right)  ^{2}\right]  \left(  R_{1}^{2}\right)  ^{2}}{(1-\xi)^{2}}
+2R_{2}^{2}R_{1}^{2}\frac{R_{1}^{1}R_{2}^{1}+R_{1}^{3}R_{2}^{3}}{1-\xi^{2}
}\right\}  \,. \label{a1-a2}
\end{equation}
According to Eq. (\ref{Q1-4-def}) we have
\begin{equation}
\frac{1}{2}\left[  \left(  R_{k}^{1}\right)  ^{2}+\left(  R_{k}^{3}\right)
^{2}\right]  =q(x_{k})+\Delta_{L}q(x_{k})=\left(  r_{k}^{+}\right)  ^{2}\,,
\end{equation}
\begin{equation}
\left(  R_{k}^{2}\right)  ^{2}=q(x_{k})-\Delta_{L}q(x_{k})=\left(  r_{k}
^{-}\right)  ^{2}\,,
\end{equation}
\begin{equation}
R_{1}^{1}R_{2}^{1}+R_{1}^{3}R_{2}^{3}\leq\sqrt{\left[  \left(  R_{1}
^{1}\right)  ^{2}+\left(  R_{1}^{3}\right)  ^{2}\right]  \left[  \left(
R_{2}^{1}\right)  ^{2}+\left(  R_{2}^{3}\right)  ^{2}\right]  }=2r_{1}
^{+}r_{2}^{+}\,.
\end{equation}
Using these expressions, we find from Eqs. (\ref{a1-a2}) and (\ref{b-beta-def})
\begin{equation}
\left(  a_{1}\right)  ^{2}+\left(  a_{3}\right)  ^{2}\leq\frac{\nu^{2}
}{2\alpha^{2}}\left(  \frac{r_{1}^{+}r_{2}^{-}}{1+\xi}+\frac{r_{2}^{+}
r_{1}^{-}}{1-\xi}\right)  ^{2}\,,
\end{equation}
\begin{equation}
b=\nu^{2}\alpha r_{1}^{-}r_{2}^{-}\,.
\end{equation}
Now we insert these results into Eq. (\ref{E-T-ineq-step-1}) and obtain the
positivity bound (\ref{E-T-q-org}).

Note that the bound (\ref{E-tilde-T-q-strong}) on $\tilde{E}_{T}^{q}$ differs
from the bound (\ref{E-T-q-strong}) on $E_{T}^{q}$ by the replacement
\begin{equation}
\nu^{2}\alpha^{4}\left(  R_{1}^{2}R_{2}^{2}\right)  ^{2}\rightarrow\xi^{2}
\nu^{2}\alpha^{4}\left(  R_{1}^{2}R_{2}^{2}\right)  ^{2}\,.
\end{equation}
Making this change in Eq. (\ref{E-T-q-org}), we arrive at the positivity bound
(\ref{E-T-tilde-q-org}).

\end{document}